%% file: main.tex
\documentclass[10pt,journal,compsoc]{IEEEtran}

\IEEEoverridecommandlockouts

\usepackage{cite}
\usepackage{booktabs} 
\usepackage{graphicx}
\usepackage{array}
\usepackage{url}
\usepackage{fancybox}
\usepackage{multirow}
\usepackage{amsmath,amssymb,amsfonts}
\usepackage{color}
\usepackage[table,xcdraw]{xcolor}
\usepackage{fancyhdr}
\usepackage{subfig}
\usepackage{diagbox}
\usepackage{bbding}
\usepackage{placeins}
\usepackage{balance}
\usepackage{hyperref} 
\usepackage{enumitem}
\usepackage[linesnumbered, ruled,vlined]{algorithm2e}
\def\BibTeX{{\rm B\kern-.05em{\sc i\kern-.025em b}\kern-.08em
		T\kern-.1667em\lower.7ex\hbox{E}\kern-.125emX}}

\newcommand{\allTeamNum}{60}
\newcommand{\OpenReportNum}{2,766}
\newcommand{\ReportCodeNum}{1,199}
\newcommand{\selectedTeamNum}{29}
\newcommand{\allWeaknessNum}{9,154}

\newcommand{\ReportNum}{608}
\newcommand{\DAppNum}{682}
\newcommand{\averageLOC}{7,885}
\newcommand{\allSolNum}{39,904}
\newcommand{\averageSolNum}{58}

\newcommand{\SWCNum}{1,618}
\newcommand{\SWCByteNum}{888}
\newcommand{\compiledByteNum}{6,665}

\begin{document}
	
	\title{DAppSCAN: Building Large-Scale Datasets for Smart Contract Weaknesses in DApp Projects}
				
   
	\author{Zibin Zheng, Jianzhong Su, Jiachi Chen, David Lo, Zhijie Zhong and Mingxi Ye
		\IEEEcompsocitemizethanks{\IEEEcompsocthanksitem Zibin Zheng, Jiachi Chen, Jianzhong Su, Zhijie Zhong and Mingxi Ye are with School of Software Engineering, Sun Yat-sen University, China. \protect\\
			E-mail: \{zhzibin, chenjch86\}@mail.sysu.edu.cn

                 E-mail: \{sujzh3, zhongzhj3, yemx6\}@mail2.sysu.edu.cn

            \IEEEcompsocthanksitem David Lo is with the School of Information Systems, Singapore Management University, Singapore.\protect\\
			E-mail: davidlo@smu.edu.sg
			
			\IEEEcompsocthanksitem Jiachi Chen is the corresponding author.}
		\thanks{Manuscript received     ; revised   }}

	\markboth{IEEE Transactions on Software Engineering, ~Vol.50, No.6, JUNE 2024 }%
	{Shell \MakeLowercase{\textit{et al.}}: Bare Demo of IEEEtran.cls for Computer Society Journals}

	\IEEEtitleabstractindextext{%
            \input{S0-abstract}

        }

	
	\maketitle
	\IEEEdisplaynontitleabstractindextext


	\input{S1-Introduction}
	\input{S2-Background}

\input{S3-Motivation}

	\input{S4-Source}
    \input{S5-Bytecode}

    \input{S6-Tool}
	\input{S7-discussion}

	\input{S8-Related}
	\input{S9-Conclusion}

\section*{Acknowledgments}
This research / project is supported by the National Key Research and Development Program of China (2023YFB2704801), the National Natural Science Foundation of China (62032025, 62302534), and the National Research Foundation, under its Investigatorship Grant (NRF-NRFI08-2022-0002). Any opinions, findings and conclusions or recommendations expressed in this material are those of the author(s) and do not reflect the views of National Research Foundation, Singapore.
	\maketitle\maketitle
	\balance
	\bibliographystyle{IEEEtran}
	\bibliography{refs}

\end{document}

%% file: S0-abstract.tex
\begin{abstract}
The Smart Contract Weakness Classification Registry (SWC Registry) is a widely recognized list of smart contract weaknesses specific to the Ethereum platform. Despite the SWC Registry not being updated with new entries since 2020, the sustained development of smart contract analysis tools for detecting SWC-listed weaknesses highlights their ongoing significance in the field.
However, evaluating these tools has proven challenging due to the absence of a large, unbiased, real-world dataset. To address this problem, we aim to build a large-scale SWC weakness dataset from real-world DApp projects. We recruited 22 participants and spent 44 person-months analyzing {\ReportCodeNum} open-source audit reports from {\selectedTeamNum} security teams. In total, we identified {\allWeaknessNum} weaknesses and developed two distinct datasets, i.e., \textsc{DAppSCAN-Source} and \textsc{DAppSCAN-Bytecode}. The \textsc{DAppSCAN-Source} dataset comprises {\allSolNum} Solidity files, featuring {\SWCNum} SWC weaknesses sourced from {\DAppNum} real-world DApp projects. However, the Solidity files in this dataset may not be directly compilable for further analysis. To facilitate automated analysis, we developed a tool capable of automatically identifying dependency relationships within DApp projects and completing missing public libraries. Using this tool, we created \textsc{DAppSCAN-Bytecode} dataset, which consists of {\compiledByteNum} compiled smart contract with {\SWCByteNum} SWC weaknesses. Based on \textsc{DAppSCAN-Bytecode}, we conducted an empirical study to evaluate the performance of state-of-the-art smart contract weakness detection tools. The evaluation results revealed sub-par performance for these tools in terms of both effectiveness and success detection rate, indicating that future development should prioritize real-world datasets over simplistic toy contracts.
\end{abstract}

%% file: S1-Introduction.tex
\section{Introduction}
\label{Introduction}

In 2015, Ethereum~\cite{ethereum} introduced a revolutionary technology named smart contracts~\cite{solc}. Smart contracts can be regarded as Turing-complete programs deployed on the blockchain. By utilizing smart contracts, developers can easily develop their decentralized applications (DApp). DApps are immutable, self-executed, without a centralized architecture, which guarantees the transparency and trustworthiness of DApps. These features make smart contracts widely used in many areas, e.g., finance~\cite{Defi} and gaming~\cite{CryptoKitties}.  

Unfortunately, a large number of security incidents related to Ethereum smart contracts have occurred and have caused billions of dollars in financial losses~\cite{slowmist} in recent years. To increase the security of smart contracts, significant effort has been devoted to identifying and detecting security issues in smart contracts. For example, Chen et al.~\cite{chen2020defining} introduced 20 kinds of smart contract defect by analyzing online Q\&A posts. The DASP project~\cite{dasp} is a smart contract taxonomy that reports on 10 vulnerabilities. A notable blockchain security team named ConsenSys~\cite{consensys} summarized several common smart contract problems and provided a repository named the \textit{Smart Contract Weakness Classification Registry (SWC Registry~\cite{SWC})}. There are 37 kinds of weaknesses in the SWC Registry as of April 2023. Based on classified weaknesses in the SWC Registry, numerous automated analysis tools employing various technologies have been developed in recent years~\cite{chen2021maintenance}, such as those based on program analysis~\cite{torres2018osiris}, formal verification~\cite{park2018formal}, fuzzing~\cite{su2022effectively}, and machine learning methods~\cite{gao2020checking}. In general, the SWC Registry is still one of the most widely used weakness classifications and covers most of the weaknesses detected by existing analysis tools.

However, evaluating these tools is often challenging due to the scarcity of a large-scale labeled dataset. Based on our investigation of 20 academic papers, two common methods have been used to evaluate these tools: (1) manually labeling a small-scale dataset (typically comprising hundreds of contracts) and using it to evaluate the effectiveness (e.g., precision, recall)~\cite{liao2022smartdagger}, and (2) using their tools to analyze a large-scale dataset (usually containing thousands of contracts) and then manually checking the correctness of some/all contracts which are labeled positive~\cite{oyente}. 

Both of these two methods have their limitations. In the \textit{first} method, the selection of the dataset might be unfair or non-representative. Although many researchers have randomly selected and labeled hundreds of smart contracts to evaluate their tools, the majority of these contracts are toy contracts with significantly fewer lines of code than real-world DApp projects. Achieving good results on a dataset predominantly composed of toy contracts cannot guarantee the tool's effectiveness for real-world DApps. Furthermore, only a small proportion of contracts contain weaknesses among hundreds of contracts, which also impact the results. Regarding the \textit{second} method, it can only determine the false/true positive rates of proposed tools; however, it cannot evaluate the tool's false/true negatives. Additionally, although the evaluation of prior tools includes a substantial number of smart contracts, approximately 96\% of them are toy contracts that contain less than five on-chain transactions~\cite{chen2020understanding}. Consequently, the second method also has limited power to evaluate the tools considering a more realistic setting.  

Based on the above motivations, the aim of this study is to construct a dataset that contains a substantial number of vulnerable contracts from real-world DApps. In the smart contract community, there are several well-known security teams, e.g., TrailofBits\cite{tob} and Slowmist~\cite{slowmist}, which provide audit services for DApp projects. Each DApp undergoes a comprehensive audit conducted by security experts, who then provide audit reports containing detailed defeat descriptions and their locations in DApps. Although audit reports effectively enhance the security of contracts, their cost is typically high. As a result, audited DApps are generally real-world projects rather than simple toy contracts.

In this paper, we recruited 22 participants (including 7 PhD and 15 master's students) and dedicated 44 person-months to manually analyze {\ReportCodeNum} open-source audit reports provided by {\selectedTeamNum} security teams. We finally summarized {\allWeaknessNum} weaknesses among these DApps. Given that most current smart contract research focuses on a few common weakness types, e.g., \textit{Reentrancy} and \textit{Integer Overflow and Underflow}, and that the majority of them can be found in the SWC Registry, we specifically highlighted {\SWCNum} SWC weaknesses from {\DAppNum} DApps. On average, each DApp has {\averageSolNum} Solidity smart contract files ({\allSolNum} in total) with {\averageLOC} lines of code, and 66.3\% of DApps have compiler versions higher than 0.6. For each SWC weakness, our dataset provides a description, location within the source code, and the associated audit report. We call this dataset \textsc{\textbf{DAppSCAN-Source}}, as the dataset provides the SWC weakness information at the source code level.

Note that a DApp typically consists of multiple Solidity files with complex dependencies; it often relies on numerous online libraries, such as \texttt{\small Safemath} in OpenZeppelin~\cite{safemath}, which may not be included in the source code of the DApp project. Consequently, Solidity contract files in the DApp project might not be able to compile directly. To facilitate the use of the dataset, we developed a tool that can identify the dependency relationships of contracts within a DApp and insert missing library code into the contracts. Using this tool, we successfully compiled {\compiledByteNum} smart contracts, and obtained their bytecode and related ABI information~\cite{ABI}. These contracts contain {\SWCByteNum} SWC weaknesses, and we call this dataset as \textsc{\textbf{DAppSCAN-Bytecode}}.

Furthermore, based on \textsc{DAppSCAN-Bytecode}, we conducted an empirical study to evaluate the effectiveness of state-of-the-art smart contract weakness detection tools, i.e., Mythril~\cite{mythx}, Slither~\cite{feist2019slither}, Security~\cite{tsankov2018securify}, Smartian~\cite{choi2021smartian}, Sailfish~\cite{bose2022sailfish} and eTainter~\cite{ghaleb2022etainter}. We found that except for Slither that can successfully analyze 86\% of contracts, other tools can only analyze a small portion of contracts in the dataset. Besides, among the contracts that these tools can analyze, only a few weaknesses are correctly detected. The results indicate that the tools should focus more on real-world situations rather than simple toy contracts.

In addition, we found that there exist 82.3\% of non-SWC weaknesses in collected audit reports. This is because the SWC registry mainly includes the classical and general weaknesses in smart contracts, e.g., \textit{Reentrancy} and \textit{Integer Overflow}, but the majority of the reported weaknesses in audit reports are non-security issues (such as code optimization suggestions) or non-classical issues (like functional bugs specific for each DApp), which are hard to be classified with general rules. Performing manual analysis on these non-classical weaknesses requires much more labor and potentially compromises data quality. Therefore, we focus only on the SWC weaknesses in this paper. Our dataset is only suitable for evaluating tools on classical and low-level smart contract weaknesses, but not on non-classical weaknesses (e.g., function bugs). Meanwhile, non-SWC weaknesses are also necessary for research because of their significant proportion in audit reports. 

The main contributions of this paper are as follows. 

\begin{itemize} \setlength{\itemsep}{4pt}
	
    \item \textit{Two large-scale SWC weakness datasets. } In this paper, we proposed two large-scale datasets from real-world DApp projects, i.e. \textsc{DAppSCAN-Source} and \textsc{DAppSCAN-Bytecode}. The dataset could aid further research on smart contract analysis. We open the whole dataset to the public at: \url{https://github.com/InPlusLab/DAppSCAN/}.

     \item \textit{A tool to obtain compiled bytecode within DApps. } We proposed a tool that can automatically analyze dependencies between contracts, based on which we can insert missing public library code to generate the compiled smart contracts bytecode from DApp projects. 
	
	
    \item \textit{Empirical study of current tools. }  We investigate the evaluation methods and dataset used by current smart contact analysis tools. We also assess state-of-the-art tools based on our dataset. The results demonstrate that most of their evaluations were based on toy contracts with a few lines of code and older compiler versions, and they have poor performance in detecting smart contract weaknesses in real-world DApps.  
	
\end{itemize} 


The organization of the rest of this paper is as follows. In Section~\ref{sec:background}, we provide the background knowledge of DApp projects, smart contract audit reports, and SWC Registry. Then, we highlight our motivation in Section~\ref{sec:motivation}. In Section~\ref{sec:source_dataset} and~\ref{sec:bytecode_dataset}, we introduce the details of the two datasets. In Section~\ref{sec:tool_evaluation}, we conducted an empirical study to evaluate the performance of state-of-the-art smart contract tools. Then, we provide the implications and describe threats to validity in Section~\ref{sec:discussion}. In Section~\ref{sec:related}, we elaborate on the related work. Finally, we conclude the whole study and mention future work in Section~\ref{sec:conclusion}.

%% file: S2-Background.tex
\section{Background}
\label{sec:background}

\subsection{Smart Contracts and DApp Projects} 
\label{sec:background_dapp}


Smart contracts are Turing-complete programs running on the blockchain. Solidity~\cite{solidity} is one of the most well-known smart contract languages, which is an object-oriented language similar to Java and C++. It supports smart contracts that inherit code from other contracts and import code from other contract files. A DApp project always consists of multiple smart contracts and files, and its file structure is usually complicated due to the inheritance and import relationship. When developing a DApp project, developers always use frameworks (e.g., Truffle~\cite{Truffle}, Brownie~\cite{Brownie}) to enhance development efficiency. These frameworks usually support import third-party contracts/library from Github directly; the code \textit{import ``@openzeppelin/contracts/token/ERC20/ERC20.sol"} can be used to download the ERC20 contract from Openzeppelin's Github repository directly, without the use of a code in the local environment. Furthermore, a DApp usually contains two parts, i.e., smart contracts and scripts. The smart contracts in the DApp might be deployed to multiple addresses, and these can interact with the scripts written in Python, Javascript, or other languages. 


\subsection{Smart Contract Audit Reports} 
\label{sec:audit}
In recent years, frequent security incidents that resulted in numerous financial losses motivated the birth of the smart contract audit service. The audit service provides a detailed code review, provided by security experts, to ensure security and potentially improve the performance of a smart contract. As of August 2022, there are {\allTeamNum} security teams that have been recommended by Etherscan~\cite{etherscan_audits}, which are well known in blockchain ecology.

The audit process usually contains four steps~\cite{audits}. First, an automated analysis step is performed using smart contract analysis tools. However, automated analysis is usually error-prone. Second, an in-depth manual code review is performed. This step is usually carried out by at least two security experts to ensure that all weaknesses are correctly discovered. Third, the security team contacts the customer and provides recommendations for security and optimization. The customers also confirm and repair the reported weaknesses during the consulting process. Finally, a security audit report is provided, which contains the descriptions, locations, and repaired versions (if provided) of weaknesses.


The price of an audit service varies and depends on the complexity of the contracts. Since auditing a DApp usually takes a long period of time and requires a high level of professional knowledge, the fees for these services are usually expensive. 

\subsection{Smart Contract Weakness Classification (SWC) Registry} 
In software engineering, weaknesses are errors that can lead to vulnerabilities, while vulnerabilities are mistakes that hackers can use to attack projects~\cite{weakness}. The smart contract weakness classification (SWC) registry~\cite{SWC}, also known as EIP-1470~\cite{eip1470}, is a specification introduced to classify common smart contract weaknesses. Each SWC weakness is included with a title, description, ways to fix it, and a specific SWC-ID. As of April 2023, 37 kinds of weakness have been collected in the SWC Registry. Most of the common issues reported in academic works can be linked to SWC weaknesses. For example, permission-less issues~\cite{liu2022finding} are classified into two weaknesses, e.g., SWC-105 (Unprotected Ether Withdrawal) and SWC-106 (Unprotected SELFDESTRUCT Withdrawal). The SWC-ID and title can be found in Table~\ref{tab:swc}; the details of the 37 kinds of SWC weakness can be found at: https://swcregistry.io/. 

In particular, although some SWC weaknesses are either irrelevant from a security perspective or repaired in the recent Solidity compilers, they are worth researching. On the one hand, while SWC weaknesses include instances that may not pose a security threat (according to CWE~\cite{cwe}), non-security-related weaknesses still merit consideration. For example, SWC-135 (Code With No Effects) is frequently highlighted in audit reports, as addressing this problem could save costs for developers. On the other hand, although new Solidity versions address some weaknesses, there are a massive number of contracts with old Solidity versions running on Ethereum, and it is still possible for developers to deploy contracts with old versions. Thus, the SWC weaknesses being addressed by new Solidity versions (e.g., SWC-100 Function Default Visibility) still deserve attention.

%% file: S3-Motivation.tex
\section{Motivation}
\label{sec:motivation}

\begin{figure} 
	\begin{center}
		\includegraphics[width=0.45\textwidth ]{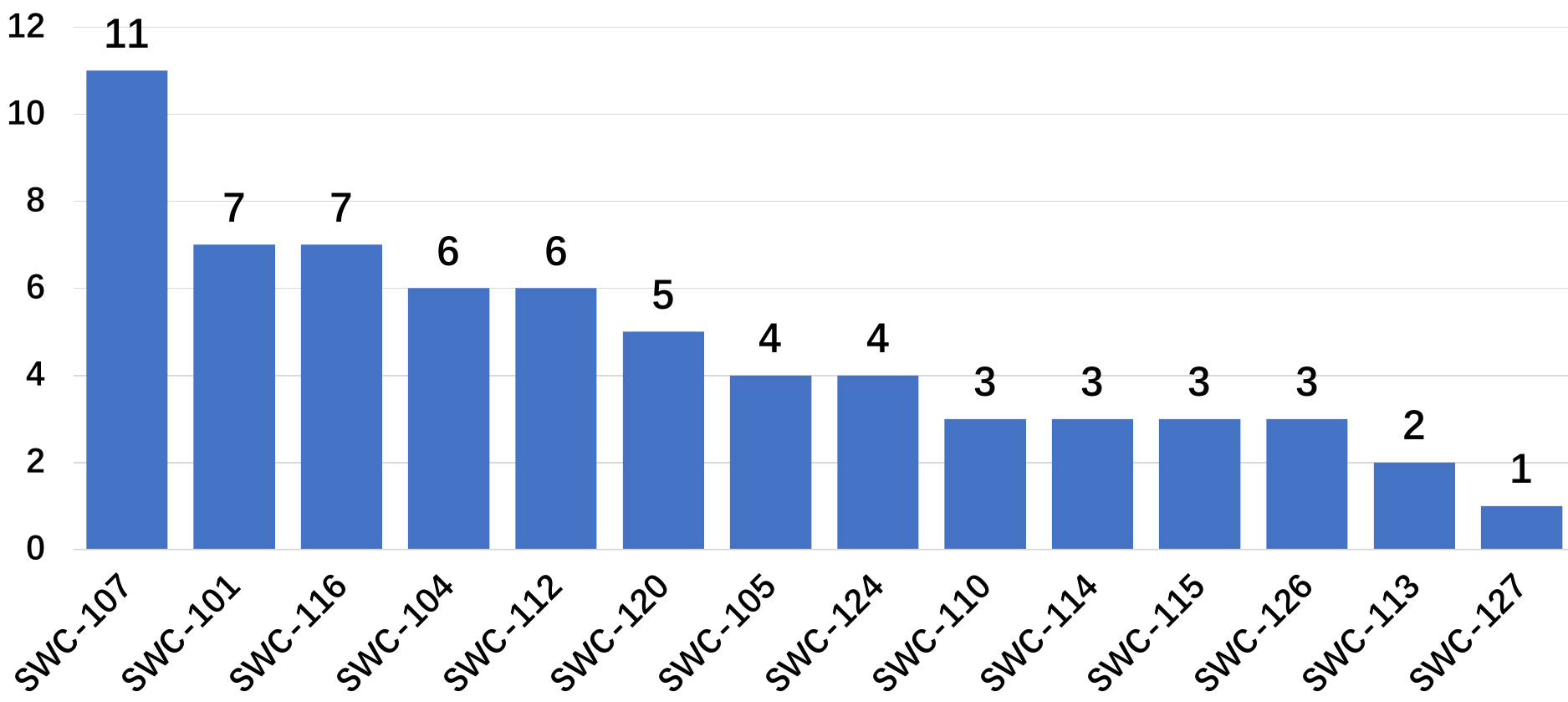} 
		\caption {The number of academic tools published in top conferences that can detect certain SWC weaknesses.} 
		\label{Fig:academia} 
	\end{center} 
\end{figure} 

In this section, we investigate the dataset and evaluation methods used by current smart contract weakness detection tools. The results indicate a high demand for a large-scale, timely, and accurate dataset, which underscores the significance of our dataset.

\subsection{Tools Collection and Analysis}

We surveyed the major smart contract weakness detection tools published at top software engineering and security conferences, i.e., ASE, FSE, ICSE, ISSTA, S\&P, CCS, USENIX Security, and NDSS. Note that a weakness may have a different name in different academic works. For example, a weakness known as ``Timestamp Dependence" was first introduced by Luu et al.~\cite{oyente}. The same weakness was also reported by sfuzz~\cite{nguyen2020sfuzz} and named as ``Block Number Dependency". This weakness is called ``Weak Sources of Randomness from Chain Attributes" in the SWC Registry. Since the SWC Registry has the largest coverage of the kinds of Ethereum weaknesses, we use the names defined by the SWC Registry in this paper. 

\begin{table*}
	\caption{20 Tools published in top SE/Security conferences that can detect SWC weaknesses.} 
	\label{tab:tools}  
	\centering
	\small
	\begin{tabular}{p{80pt}  | p{40pt}<{\centering} | p{160pt} | p{150pt}}
		\hline
		\textbf{Tools} & \textbf{Venues} & \textbf{Detected SWC ID} & \textbf{Dataset and Evaluation}\\
		\hline
		Oyente~\cite{oyente} & CCS'16& 104, 107, 114, 120 & M2\{T2(19,366)\}\\ 
		\hline 
		Zeus~\cite{kalra2018zeus} & NDSS'18& 101, 104, 107, 113, 114, 115, 120 & M2\{T1(1,524)\}\\ 
		\hline 
		teEther~\cite{krupp2018teether} & Security'18& 106, 112 & M2\{T2(38,757)\}\\ 
		\hline 
		Securify~\cite{tsankov2018securify} & CCS'18& 104, 105, 107, 112, 114, 124 & M1\{T1(100)\} + T2(24,594)\\ 
		\hline 
		ContractFuzzer~\cite{jiang2018contractfuzzer} & ASE'18& 107,  112, 116, 120, 126 & M2\{T1(6,991)\}  \\ 
		\hline 
		sFuzz~\cite{nguyen2020sfuzz} & ICSE'19& 101, 107,  112, 116, 120, 126 & M2\{T1(4,112)\}\\ 
		\hline 
		Sereum~\cite{rodler2018sereum} & NDSS'19& 107 & M2\{T2(24,594 )\}\\ 
		\hline 
		Smartian~\cite{choi2021smartian} & ASE'19& 101, 104, 105, 106, 107, 115, 116, 124, 127 & M3\{T1(58)\} + M3\{T1(72)\}  + T3(500)\\ 
		\hline 
		ILF~\cite{he2019learning} & CCS'19& 104, 105, 106, 112, 116 & M2\{T1(18,496)\}\\ 
		\hline 
		Harvey~\cite{wustholz2020harvey} & FSE'20& 110, 124 & M2\{T1(17)\}\\ 
		\hline 
		extended Harvey~\cite{wustholz2020targeted} & ICSE'20& 110, 124 & M3\{T1(17)\}\\ 
		\hline 
		CLAIRVOYANCE~\cite{xue2020cross} & ASE'20& 107 & M2\{T1(17,770)\}\\ 
		\hline 
		SOLAR~\cite{feng2020summary} & ASE'20& 107, 116, 120 , 126 & M2\{T1(25,983)\}\\ 
		\hline 
		VeriSmart~\cite{so2020verismart} & S\&P'20& 101 & M3\{T1(60)\}\\ 
		\hline 
		VerX~\cite{permenev2020verx} & S\&P'20& 101, 110 & M3\{T1(12)\}\\ 
		\hline 
		ETHBMC~\cite{frank2020ethbmc} & Security'20&106, 112 & M5\{T2(2,194,650)\}\\ 
		\hline 
		SmartTest~\cite{so2021smartest} & Security'21& 101, 105, 106 & M4\{T1(443)\}\\ 
		\hline 
		Unknown Name~\cite{ren2021making} & FSE'21& 101, 104, 107, 115,  116 & M3\{T1(176)\} + T1(47,398) \\
		\hline 
		Smartdagger~\cite{liao2022smartdagger} & ISSTA'22& 107,  113, 116 & M5\{T1(47,398)\} + M1\{T1(594)\}\\ 
		\hline
		eTainter~\cite{ghaleb2022etainter} & ISSTA'22 & 113, 128 &  M1\{T1(28)\} + T2(60,612) + T3 (3,000)\\
		\hline
	\end{tabular} 
\end{table*}

\begin{table}
	\caption{Types and methods used to evaluate tools}
	\label{tab:evaluation}  
	\centering
	\small
	\begin{tabular}{l | p{210pt}}
		\hline
		\textbf{ID} & \textbf{Description} \\
		\hline
		T1 & Verified smart contracts collected from Etherscan \\
		\hline
		T2 & Smart contract on Ethereum without the need for source code\\
		\hline
	   T3  & A selected smart contract dataset with high transactions or balances\\
		\hline
		T4 & DApp projects collected from Github \\
		\hline
		M1 & Manually label the vulnerabilities in a dataset before testing the tool \\
		\hline
		M2 & Manually check all/some of the cases detected by the tool. \\
		\hline
		M3 & Reuse previously labeled datasets \\
		\hline
		M4 & Reuse contracts from CVE (https://www.cve.org/) \\
		\hline
		M5  & Compare the performance with previous tools \\	
		\hline
	\end{tabular} 
\end{table} 

\subsection{Results}

We totally find 20 tools published in top SE/Security conferences at the time we conducted our study. Table~\ref{tab:tools} lists their names, the venues in which they were published, the SWC weaknesses that can be detected using these tools, and the dataset and evaluation method they use. The number of tools that can detect certain SWC weaknesses is also given in Figure~\ref{Fig:academia}. From the figure, we can find that only 14 out of 37 SWC weaknesses could be detected using the current tools. \textit{Reentrancy} (SWC-107, 11 times) is the most popular weakness investigated in academia, followed by \textit{Integer Overflow and Underflow} (SWC-101, 7 times) and \textit{Block Values as a Proxy for Time} (SWC-116, 7 times). Most of the SWC weaknesses are currently not supported by these tools. 


The datasets and evaluation methods used in these tools are listed in the last column of Table~\ref{tab:tools}. There were four types of dataset (T1 to T4) and five kinds of method (M1 to M5) used to evaluate tools, which are listed in Table~\ref{tab:evaluation}. Each evaluation method is represented in the form of ``Method\{Type(Size of Dataset)\}". For example,  M2\{T1(1,524)\} means that the dataset has 1,524 verified smart contracts (T1)~\footnote{T1 means the first type of the dataset}. Then, the authors manually checked some/all the cases detected by the tool (M2). The symbol ``+" means that multiple evaluations were conducted. For example, M3\{T1(176)\} + T1(47,398) means that a previous labeled dataset (M3), which consists of 176 verified smart contracts (T1), was reused to evaluate the tool. Then, the tool was used to analyze 47,398 smart contracts (T1) without checking their correctness. 

Only two works (Smartain~\cite{choi2021smartian} and eTainter~\cite{ghaleb2022etainter}) used a selected dataset with high-value transactions or balances, which means that most of the works used toy contracts to evaluate their tools. For example, the dataset of SmartBug was reused by tools published in the past two years~\cite{ren2021making, liao2022smartdagger, choi2021smartian}. However, this dataset may include many out-of-date and toy contracts, as the average number of lines of code for this dataset is 204, and 99.8\% of the contracts' compiler versions are lower than 0.6, while the latest Solidity compiler version is 0.8+.

Moreover, M2 was widely employed to evaluate the tools, though it could only verify the false/true positive results of a contract without reporting false/true negative results. While M1 can overcome this limitation, its dataset size is typically small. These limitations underscore the necessity for a dataset comprising a substantial number of labeled weakness cases. 


%% file: S4-Source.tex
\section{The DAppSCAN-Source Dataset}
\label{sec:source_dataset}

The previous section highlighted the need for a dataset with: (1) a reasonable number of weakness cases, and (2) timely contracts from real-world DApp projects rather than toy contracts with old Solidity versions. In this section, we present the methodology employed to construct such a dataset. 

\subsection{Methodology}
\begin{figure} 
	\begin{center}
		\includegraphics[width=0.45\textwidth]{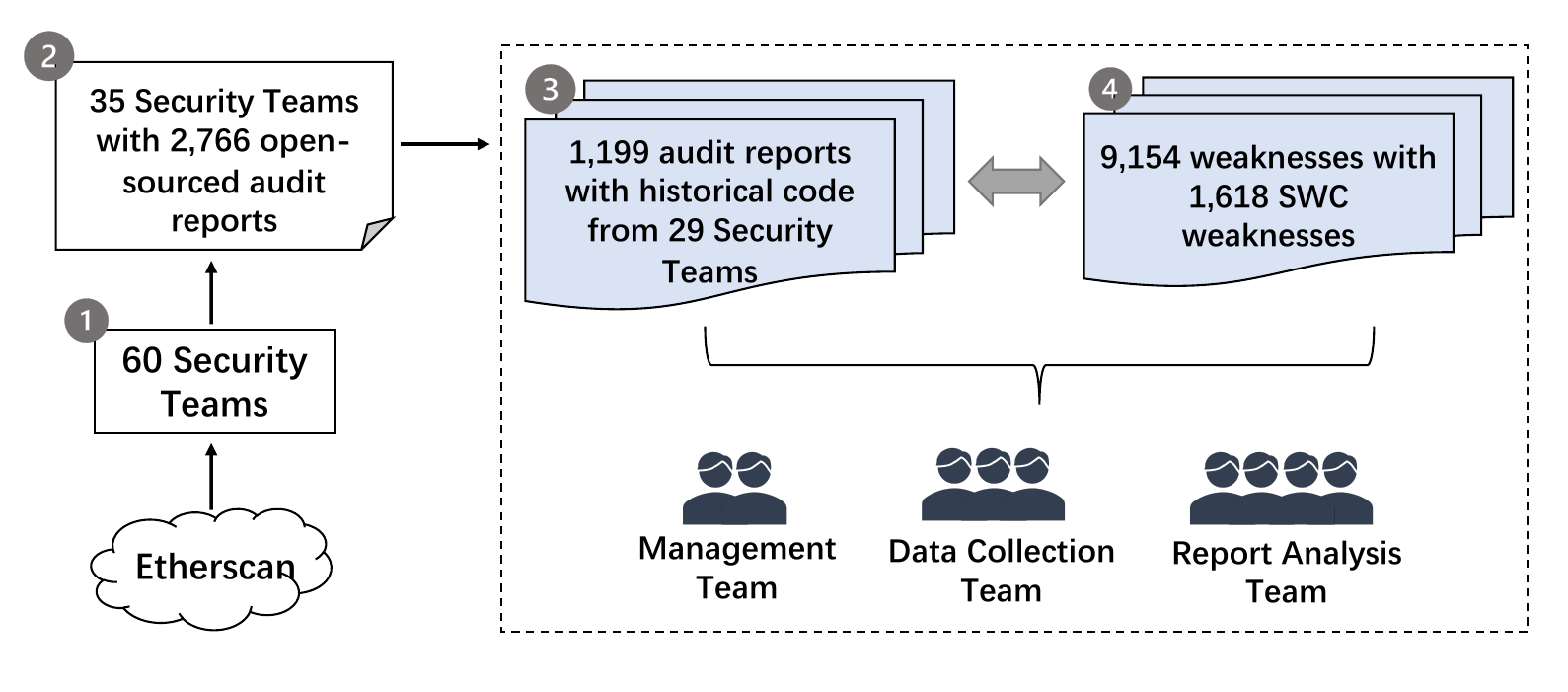} 
		\caption {Overview architecture of finding SWC weaknesses from audit reports.}  
		\label{Fig:swc} 
	\end{center} 
\end{figure}

Audit reports are documents that contain detailed security information about smart contracts. Figure~\ref{Fig:swc} illustrates the process used to find SWC weaknesses in DApp audit reports. First, we obtained a list of {\allTeamNum} recommended security teams from Etherscan (as of August 2022) and found that there are only 35 teams that provide open-source audit reports with permissions for content extraction and redistribution. Then we manually reviewed the information on their homepages and discovered {\OpenReportNum} open-source audit reports. Then, we include the audit reports that provide historical source code. After that, we include {\ReportCodeNum} audit reports from {\selectedTeamNum} security teams. Finally, we manually analyzed the selected audit reports and found {\allWeaknessNum} weaknesses; {\SWCNum} of them are SWC weaknesses. These SWC weaknesses were derived from {\DAppNum} DApps and {\ReportNum} audit reports (we only focused on Solidity projects). In the following, we introduce the details of each step.

\subsubsection{Participant Recruitment}
\label{sec:swc_recruit}
Analyzing such a large number of audit reports is a labor-intensive task. To alleviate the workload, we need to assemble a team of participants familiar with Solidity program and SWC weaknesses. The first author's university has a blockchain laboratory with experienced researchers. Therefore, we sent invitations to the researchers in the laboratory to introduce the details of our work and the process of analyzing audit reports. We finally recruited 22 researchers for this task.


\subsubsection{Roles of Team Members}
Although all the recruited researchers claimed that they have rich experience in blockchain area, they may have had different understandings of SWC weaknesses and Solidity programming. Thus, we interviewed each member to understand their background. Based on these interviews, we found that three researchers had professional knowledge of smart contract weaknesses; they have already published several smart-contract-related works at top venues. Also, five researchers had limited background in Solidity weaknesses, but all had good experience in programming and blockchain. Regarding the other 14 researchers, they had good knowledge in smart contract programming but had not yet published related papers in top conferences/journals. On the basis of their background, we divided them into three groups. 

\begin{itemize} \setlength{\itemsep}{2pt}
    \item \textbf{Management team.} This team consisted of the three researchers with professional knowledge. Their duties included (1) launching online training on SWC weaknesses, (2) answering other members' questions, (3) reviewing audit report analysis results and refining the data, and (4) managing the work schedule. 
	
    \item \textbf{Data collection team.} This team consisted of five researchers with a limited background in Solidity weaknesses. Their duties are described as step 2 and step 3 in Figure~\ref{Fig:swc}, namely, (1) finding out how many security teams released open-source audit reports and downloading all the audit reports; (2) reviewing the collected audit reports and locating the historical DApp versions containing SWC weaknesses as described in the audit reports, and (3) removing non-Solidity DApp projects, as our focus is Solidity-based DApps.
	
    \item \textbf{Report analysis team.} The remaining 14 members responded to analyzing the audit reports. They were required to (1) summarize the weaknesses described in the audit reports, (2) determine whether the weaknesses belonged to the SWC registry, and (3) if so, provide the SWC-ID and pinpoint its location within the source code.
	
\end{itemize}		

\subsubsection{Data Collection}
The data collection team was responsible for collecting two types of data, i.e. audit reports and the codes of DApps. To collect audit reports, we examined the homepages of the security teams, as they typically provide a list of reports on their websites. Then we saved all of the audit reports as PDFs to avoid potential future link unavailability. In total, we collected {\OpenReportNum} open-source audit reports from 35 security teams that release their audit reports with permissions for content extraction and redistribution. To ensure data completeness, two team members analyzed the reports of each security team.

For DApp codes, most weaknesses described in the audit reports had already been patched in the latest versions. Fortunately, audit reports usually highlight the code version they analyzed, e.g., a historical github version or a link on Etherscan/BscScan~\cite{bsc}. Thus, we manually reviewed the audit reports to find the corresponding code versions. As a result, we found {\ReportCodeNum} audit reports specifying the audited code version, which belong to {\selectedTeamNum} security teams. For each DApp in these audit reports, we also saved its code to prevent potential data loss and perform a double check to ensure correctness.

\subsubsection{Data Analysis}
\label{sec:swc_analysis}
This step was carried out by the report analysis team and was assisted by the management team. Before analyzing audit reports, 14 members of the report analysis team received training from the management team. The training content included: (1) information about the SWC Registry and examples of each SWC weakness, (2) how to analyze an audit report, and (3) the content of the analysis report they should submit. An analysis report contained a summary of the weakness, the type of SWC, and the location of the weakness in the source code. After training, {\ReportCodeNum} audit reports were distributed to the report analysis team in four rounds. 


\begin{itemize} \setlength{\itemsep}{2pt}
    \item \textbf{First Round.} In the first round, 188 out of {\ReportCodeNum} (15.7\%) audit reports were distributed to the report analysis team (almost 27 reports per member), and each audit report was analyzed by two members independently. To ensure independence, the analysis team members were not allowed to share their results and dataset with others. The results were then integrated by the management team. Then, the management team compared and checked the results to summarize some common issues, e.g., formatting errors and misunderstandings of SWC weaknesses. Next, the report analysis team received the second stage of training on these common issues. After training, the members were required to self-check their results and compare their results with another member. If the result was different, they were required to discuss it together and provide a final result. The management team provided technical help throughout the process. 
	
    \item \textbf{Second and Third Rounds.} In the second and third rounds, 163 (13.6\%) and 284 (23.7\%) audit reports were distributed to the report analysis team, respectively. Similarly to the first round, each audit report was also independently analyzed by two members. Then, the results were collected by the management team and compared for consistency. The difference in the results would lead to the results being returned, and the report analysis team members would be required to discuss the difference and further provide a consistent result. If they were unable to reach an agreement, the management team would help to make a final decision. 
	
    \item \textbf{Fourth Round.} All the remaining audit reports were distributed in this round. After the second and third rounds, we found that the correctness rate increased rapidly with the progress of the project. Thus, in the fourth round, we did not double check. Instead, the management team randomly sampled and checked 10\% of the analysis results. The sampling results showed the good quality of the analysis. 
	
\end{itemize}	

\subsection{Results}
\label{subsec:SWC-result}

\subsubsection{Dataset}

\begin{table}
	\caption{Statistical information of \textsc{DAppScan-Source}.} 
	\label{tab:dapp}  
	\centering
	\small
	\begin{tabular}{l | c}
		\hline
		\textbf{Key Information} & \textbf{Numbers} \\
		\hline
            Total audit reports & {\ReportNum} \\
            \hline
		Total DApps& {\DAppNum} \\
            Total Solidity files& {\allSolNum} \\
            Average Solidity files in a DApp & {\averageSolNum} \\
		Average Line of Code in a DApp& {\averageLOC} \\
		\hline
		Compiler Version 0.4+ & 104 \\
		Compiler Version 0.5+ & 123 \\
		Compiler Version 0.6+ & 164 \\
		Compiler Version 0.7+ & 48 \\
		Compiler Version 0.8+ & 240 \\
		Other Compiler Version  & 3 \\
		\hline
		
	\end{tabular} 
\end{table}

Through manual data analysis of audit reports, we totally found {\allWeaknessNum} weaknesses. In this work, we focus on SWC weaknesses and build up \textsc{DAppScan-Source} dataset based on the audit reports containing at least one SWC weaknesses. There are three parts of the \textsc{DAppScan-Source} dataset, i.e., audit reports, the codes of DApps, and analysis reports. 

{\bf Audit Reports.} Overall, we found SWC weaknesses in {\ReportNum} audit reports, which have the corresponding codes of DApps.  Note that some audit reports provide the codes before and after the revision, which are both collected to facilitate follow-up research.

{\bf DApps.} Since an audit report might review multiple DApps, we totally found {\DAppNum} DApps from the {\ReportNum} audit reports. Table~\ref{tab:dapp} lists some key information about DApps. Each DApp has about {\averageSolNum} Solidity smart contract files ({\allSolNum} in total) with {\averageLOC} lines of code on average. 
In general, \textbf{the real-world DApp projects in our dataset are much more complicated than the smart contracts analyzed by previous academic tools.} For example, the average line of code in the widely used SmartBug dataset~\cite{durieux2020empirical} is only 204, and more than 90\% of their contracts are used out-of-date compiler versions (0.4+). While Solidity version 0.8+ was the most common version, about 35.2\% ($\frac{240}{682}$) in our dataset, and it was also the latest compiler version as of May 2023. Version 0.6.1 was published on Jan. 2020, and it was also a widely used version in real-world smart contract development at the time of conducting this study; about \textbf{66.3\%} ($\frac{452}{682}$) of DApps' compiler versions in our dataset were higher than 0.6, while the corresponding number on SmartBug dataset was only \textbf{0.02\%}. (The distribution of compiler versions in the SmartBug Dataset was: 0.4+: 43,664 (93.2\%); 0.5+: 3,147 (6.7\%); 0.6+: 11 (0.02\%); others: 25 (0.05\%)~\cite{durieux2020empirical}.)

\begin{figure}
    \centering
    \includegraphics[width=\linewidth]{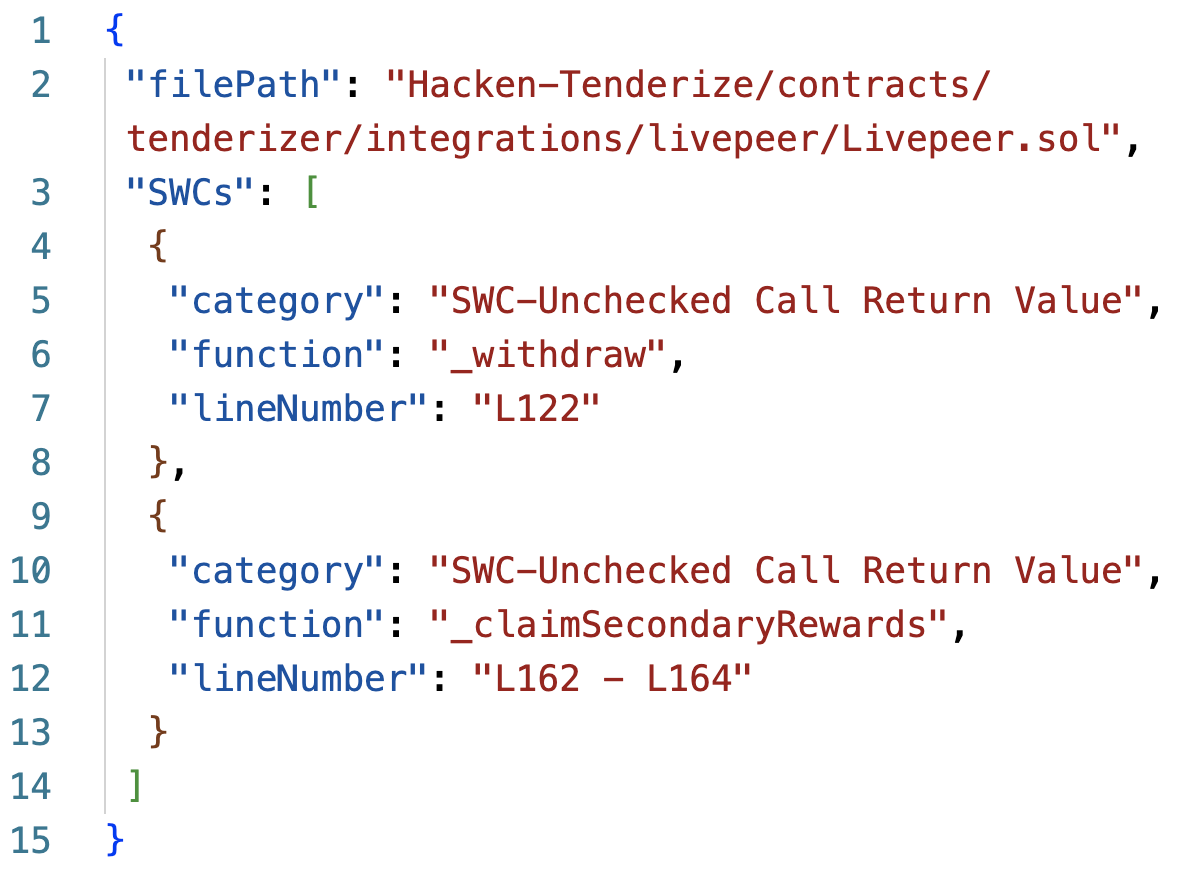}
    \caption{A sample of the analysis report in JSON format.}
    \label{fig:JSONsample}
\end{figure}

{\bf Analysis Reports.} Based on the collected codes of DApps, we convert audit reports to our designed analysis reports, which are more suitable for automated analysis. Analysis reports contain a set of JSON files that record the analysis results. Each JSON file corresponds to a Solidity file and records the SWC weaknesses in it. Fig.~\ref{fig:JSONsample} shows an example of the content of the JSON data. The attribute ``filePath'' records the path of the corresponding Solidity file, while the attribute ``SWCs'' contains the descriptions of the SWC weaknesses in the smart contract. There are three elements in each SWC weakness, that is, the SWC's \textit{category}, its corresponding \textit{function} (if exists), and the \textit{lineNumber}. The \textit{category} provides the type of SWC weakness. The \textit{function} provides the function name to which the weakness code belongs. However, some SWC weaknesses do not correspond to a function, but to a whole contract or Solidity files. For example, the ``Outdated Compiler Version" weakness can be related to all Solidity files of the DApp. In such cases, we set the \textit{function} as $N/A$ in the JSON. The \textit{lineNumber} provides the detailed location of the SWC weakness in the Solidity file, which may include multiple lines of codes or even the whole Solidity file.

\begin{table}
 \caption{The number of SWC weaknesses in two datasets. }
 \label{tab:swc}  
	\centering
	\small
	\begin{tabular}{p{15pt} <{\centering} | p{135pt}| p{30pt}<{\centering}| p{30pt}<{\centering}}
		\hline
		\textbf{ID} & \textbf{Title}  & \textbf{\# SWC in D1} & \textbf{\# SWC in D2}\\
		\hline
\textit{135} & Code With No Effects & 284 & 161 \\
\hline
\textit{101} & Integer Overflow and Underflow & 203 & 123 \\
\hline
\textit{107} & Reentrancy & 138 & 86 \\
\hline
\textit{104} & Unchecked Call Return Value & 118 & 201 \\
\hline
\textit{102} & Outdated Compiler Version & 115 & 20 \\
\hline
\textit{103} & Floating Pragma & 105 & 90 \\
\hline
\textit{128} & DoS With Block Gas Limit & 103 & 49 \\
\hline
\textit{114} & Transaction Order Dependence & 87 & 47 \\
\hline
\textit{100} & Function Default Visibility & 74 & 8 \\
\hline
\textit{131} & Presence of unused variables & 51 & 46 \\
\hline
\textit{116} & Block values as a proxy for time & 48 & 3 \\
\hline
\textit{105} & Unprotected Ether Withdrawal & 42 & 15 \\
\hline
\textit{108} & State Variable Default Visibility & 36 & 6 \\
\hline
\textit{119} & Shadowing State Variables & 34 & 0 \\
\hline
\textit{113} & DoS with Failed Call & 29 & 7 \\
\hline
\textit{129} & Typographical Error & 23 & 0 \\
\hline
\textit{120} & Weak Sources of Randomness from Chain Attributes & 23 & 5 \\
\hline
\textit{123} & Requirement Violation & 14 & 0 \\
\hline
\textit{112} & Delegatecall to Untrusted Callee & 12 & 10 \\
\hline
\textit{134} & Message call with hardcoded gas amount & 9 & 0 \\
\hline
\textit{126} & Insufficient Gas Griefing & 9 & 1 \\
\hline
\textit{124} & Write to Arbitrary Storage Location & 9 & 0 \\
\hline
\textit{111} & Use of Deprecated Solidity Functions & 9 & 0 \\
\hline
\textit{115} & Authorization through tx.origin & 8 & 1 \\
\hline
\textit{110} & Assert Violation & 7 & 0 \\
\hline
\textit{122} & Lack of Proper Signature Verification & 5 & 2 \\
\hline
\textit{125} & Incorrect Inheritance Order & 4 & 1 \\
\hline
\textit{117} & Signature Malleability & 4 & 1 \\
\hline
\textit{133} & Hash Collisions With Multiple Variable Length Arguments & 3 & 2 \\
\hline
\textit{121} & Missing Protection against Signature Replay Attacks & 3 & 1 \\
\hline
\textit{118} & Incorrect Constructor Name & 3 & 0 \\
\hline
\textit{106} & Unprotected SELFDESTRUCT Instruction & 3 & 1 \\
\hline
\textit{132} & Unexpected Ether balance & 2 & 0 \\
\hline
\textit{109} & Uninitialized Storage Pointer & 1 & 1 \\
\hline
\textit{136} & Unencrypted Private Data On-Chain & 0 & 0 \\
\hline
\textit{130} & Right-To-Left-Override control character (U+202E) & 0 & 0 \\
\hline
\textit{127} & Arbitrary Jump with Function Type Variable & 0 & 0 \\
\hline
		/ & Total & {\SWCNum} & {\SWCByteNum} \\
		\hline
	\end{tabular}  
\end{table}

\subsubsection{SWC in DApps}
\label{sec:dapp}
Totally, there are {\SWCNum} SWC weaknesses in \textsc{DAppScan-Source} that cover 34 (out of 37) kinds of SWC weakness. Among them, eight kinds of SWC weakness commonly appear in DApps (more than 100 times). The third column (D1) of Table~\ref{tab:swc} presents the number of SWC weaknesses included in \textsc{DAppSCAN-Source} dataset.

SWC-135 (Code With No Effects) was the most common SWC weakness identified. This weakness typically results from redundant code and does not cause severe security issues. However, it reduces code readability and increases gas consumption.

SWC-101 (Integer Overflow and Underflow) and SWC-107 (Reentrancy) ranked as the second and third most common weaknesses. These two weaknesses are notorious in Ethereum since they have caused significant financial losses. For example, the well-known ``The DAO" security incident~\cite{DAO} is caused by Reentrancy. The high frequency of these two SWC weaknesses highlights the seriousness of smart contract security issues.

For 15 SWC weaknesses, we found fewer than 10 cases in our dataset; many of them had already been addressed in the latest Solidity compiler or IDE can generate warnings that prevent developers from making these mistakes. For instance, SWC-118 (Incorrect Constructor Name) can be regarded as a typographical error in the code. Before Solidity version 0.4.22, the name of the constructor function had to be the same as the contract name. Thus, a typographical error in the constructor function could cause serious security problems. For example, a contract could be named \textit{BuyToken} but the constructor's name could be \textit{buyToken}, in this case, the attacker could invoke \textit{buyToken} to perform malicious behavior (e.g., modifying essential state variables). Furthermore, the introduction of SWC-111 (Deprecated Solidity Functions) and SWC-109 (Uninitialized Storage Point) weaknesses will also occur a warning in Solidity IDE, e.g., the newest version of Remix~\cite{remix}. Thus, using the latest version of the Solidity compiler and IDE can increase the security of contracts. 

In addition, \textsc{DAppScan-Source} does not contain three uncommon kinds of SWC weakness, SWC-136 (Unencrypted Private Data On-Chain), SWC-130 (Right-To-Left-Override control character (U+20E)), and SWC-127 (Arbitrary Jump with Function Type Variable).


%% file: S5-Bytecode.tex
\section{The DAppSCAN-Bytecode Dataset}
\label{sec:bytecode_dataset}



\begin{figure*} 
	\begin{center}
		\includegraphics[width=0.95\textwidth]{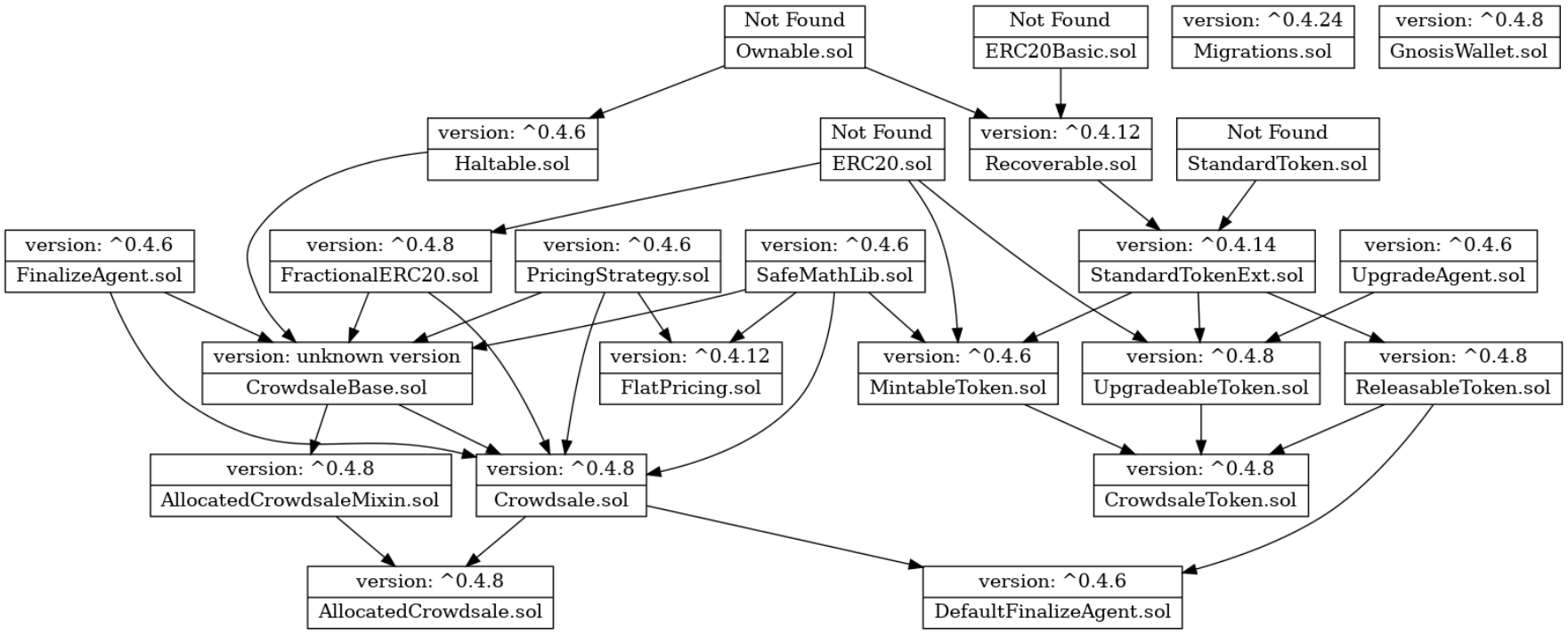} 
		\caption {Dependency relations of a DApp sample.}  
		\label{fig:drg} 
	\end{center} 
\end{figure*} 

In the previous section, we obtained a smart contract dataset directly from DApp projects, which provides the real conditions of SWC weaknesses in audit reports for real-world DApps, such as the distribution of weaknesses. In particular, \textsc{DAppSCAN-Source} is suitable for evaluating tools that do not require compilation or testing, such as emerging large-language models~\cite{chen2023chatgpt}. However, due to the complexity of DApps, directly analyzing DApps can be challenging for most existing tools since they rely on contract compilation or testing. As shown in Figure~\ref{fig:drg}, DApps typically consist of multiple contract files with complex dependencies. On the one hand, most existing tools cannot directly analyze the whole DApp project since they cannot find the main contacts of the DApp. On the other hand, the DApp relies on a significant number of public libraries (e.g., \texttt{\small SafeMath} in OpenZeppelin~\cite{safemath}), which are not included in the original DApp projects and impede compilation.

To this end, we aim to build a compilable dataset to facilitate the evaluation of existing tools. We propose a tool for compiling DApps that can identify dependency relationships and compile any contracts within DApps. In addition, we have manually collected a list of public libraries that each DApp relies on, based on which we perform automatic compilation and further build the \textsc{DAppSCAN-Bytecode} dataset.


\subsection{Tools to Compile DApps}

We present a tool for parsing the dependency relations of DApps and performing compilation. This tool comprises three critical steps, namely dependency analysis, library collection, and contract compilation.

\subsubsection{Dependency Analysis}

To better understand the structure of DApps, we first automatically identify the dependency relations in DApps to help users gain a better understanding of them and identify missing library contracts. Specifically, we scan all Solidity files in the DApp project, excluding irrelevant configuration files. For each Solidity file, we identify its dependencies based on the corresponding source code and construct a directed acyclic graph representing the relations of the DApp (dependency graph). Furthermore, each node with no out-degree in the dependency graph represents a \textit{main contract} to be compiled and deployed by DApp developers. Although we are capable of analyzing all contracts in the DApp, the following procedures are based on the leaf nodes since other contracts are redundant and overlap with them.

\subsubsection{Library Collection}

With the dependency relations of each DApp, we can identify missing library contracts such as ERC20 in OpenZeppelin. We manually collect these missing library contracts from Github and automatically insert them into the DApp project. Specifically, we identify the source of each missing library contract based on its library name (i.e. root path) and the corresponding contract path to determine its version. Different versions of the same library usually have a distinct structure. Additionally, we save these library contracts in the same directory as the DApp project.

\subsubsection{Contract Compiling}

With the structure of the DApps constructed and the missing dependencies filled, we can compile the main contracts in the DApps. This compiling procedure consists of two steps, including the identification of the Solidity version and the generation of an output file (a JSON file) that contains the compiled result. In the first step, we parse the main contract and identify its Solidity version. In particular, we omit identifying the Solidity version of its dependencies, assuming that their versions are the same as the main contract. This is because the inconsistency of the Solidity version among contracts would make compilation fail, and we do not include these cases in our dataset. After identification, we use the corresponding Solidity compiler to compile the main contracts and obtain their bytecode and ABI information.

\subsection{Results}

We have successfully compiled {\compiledByteNum} smart contracts from {\DAppNum} DApps, with {\SWCByteNum} SWC weaknesses as \textsc{DAppSCAN-Bytecode} dataset. The distribution of these SWC weaknesses is illustrated in the fourth row (D2) of Table~\ref{tab:swc}. 25 out of 37 SWC weaknesses can be found in this dataset, and eight SWC weaknesses appeared more than 40 times.

This compilable dataset contains notorious SWC weaknesses on Ethereum that can lead to huge financial losses, including SWC-104 (Unchecked Call Return Value), SWC-101 (Integer Overflow and Underflow), and SWC-107 (Reentrancy). Additionally, the high frequency of these SWC weaknesses shows the seriousness of smart contract security issues. Our dataset can assist future research in achieving a more realistic evaluation of the tools that claim to detect these critical weaknesses.

Some weaknesses only contain less than 10 cases in our compilable dataset due to: (1) the sparseness of weaknesses in real-world contracts, as we discuss in Section~\ref{subsec:SWC-result}; (2) the compilation errors caused by incomplete DApp project. Specifically, the compilation of fails mainly due to two reasons. The first reason is that some of the contracts in the project are missing, which leads to compilation failure of contracts inherited from the missing ones. The second reason is the misconfiguration of the contract version. The mismatch between the contract being compiled and the contract being inherited leads to a compilation failure.

Note that the bytecode compiled from a DApp project might contain the same weakness. For example, there are three Solidity contracts in a DApp, namely \textit{A}, \textit{B}, and \textit{C}; contract \textit{A} has a Reentrancy issue. Both \textit{A+B} (B inheriting from A) and \textit{A+C} (C inheriting from A) are compilable and can generate a bytecode with Reentrancy. This is also the reason why the number of SWC-104 in \textsc{DAppSCAN-Bytecode} is larger than that in \textsc{DAppSCAN-Source} in Table~\ref{tab:swc}. We do not remove these cases, as their bytecode is different, and both of them may exist in real-world scenarios.




%% file: S6-Tool.tex
\section{Tools Evaluation}
\label{sec:tool_evaluation}

In this section, we selected state-of-the-art weakness detection tools and use them to conduct experiments on the \textsc{DAppSCAN-Bytecode} dataset. To this end, we evaluated the effectiveness of the tools on weakness detection.

\subsection{Tool Selection}

To carry out our study, we selected representative tools for weakness detection. Although many tools have been proposed in recent years, not all of them are suitable for our study. Therefore, we define several criteria to filter appropriate tools:

\begin{itemize}
\item \textbf{State-of-the-art.} The tools are published in top-tier software engineering and security venues or well known in industry (e.g., having thousands of stars on GitHub) before May 2023.

\item \textbf{Avaliable and scalable.} The tools are publicly available and convenient for analyzing large-scale smart contracts (e.g., supporting a command-line interface).

\item \textbf{Supporting multiple Solidity versions.} Since the smart contracts in our dataset are of multiple Solidity versions, the tools should have the ability to analyze the smart contracts in various Solidity versions. 

\item \textbf{Only requiring code as input.} The input of tools only requires the Solidity code, ABI or Bytecode of smart contracts. In other words, we exclude the tools that require users to provide manual designed rules (e.g., invariants or specification) for automated analysis.

\end{itemize}

Based on above-mentioned criteria, we selected seven representative weakness detection tools whose techniques are varied, including symbolic execution, formal verification, and fuzzing. The tools are listed in~\ref{tab:selected_tools} and are briefly described as follows.

\textbf{Mythril}~\cite{mueller2018smashing} is a symbolic executor that combines taint analysis and control flow check to accurately detect smart contract vulnerabilities. In particular, Mythril has been packaged as an industry product by Consensys~\cite{consensys}.

\textbf{Securify}~\cite{tsankov2018securify} is a vulnerability detection tool based on Datalog analysis. It first abstracts the dependency graph from smart contracts, then verifies the pre-defined compliance/violation patterns to prove whether a property or vulnerability holds or not. 

\textbf{Slither}~\cite{feist2019slither} is a Solidity static analysis framework, which first converts Solidity smart contracts into an intermediate representation (IR) called SlithIR and then further verifies the contract properties for vulnerability detection and other in-depth analysis. In particular, Slither is a well-known industry tool with over 3.9k stars on GitHub.

\textbf{Smartian}~\cite{choi2021smartian} detects smart contract weaknesses based on fuzzing. It first statically analyzes the contract to initialize high-quality seed corpus. Based on the seeds, Smartian performs a lightweight dynamic data-flow analysis during testing to select more effective seeds to improve efficiency. 

\textbf{Sailfish}~\cite{bose2022sailfish} detects state-inconsistency bugs such as SWC-107 (Reentrancy) and SWC-114 (Transaction order dependence) in smart contracts. It first converts the contract into a storage dependency graph, queries it to explore vulnerability, and then uses symbolic evaluation to prune potential false alarms.

\textbf{eTainter}~\cite{ghaleb2022etainter} utilizes static taint tracking to detect gas-related vulnerabilities (e.g., SWC-128 (DoS With Block Gas Limit)) in the bytecode of smart contracts. It tracks taints through the contract state variable and the data flow from contract entry points, while using domain-specific optimizations to improve precision.

\textbf{Echidna}~\cite{grieco2020echidna} is designed for fuzzing/property-based testing of smart contracts. It uses sophisticated grammar-based fuzzing campaigns based on a contract ABI to falsify user-defined predicates or Solidity assertions. In particular, Echidna also provides built-in predicates to detect some bugs, such as SWC-101 (Integer Overflow and Underflow).

\begin{table}[t] 
    \centering
        \caption{Seven representative vulnerability detection tools used in our study.}
        \resizebox{\linewidth}{!}{
            \begin{tabular}{l|l|l}
                \hline
                \textbf{Tool} & \textbf{GitHub Repository} & \textbf{Docker Image} \\
                \hline
                Mythril & github.com/ConsenSys/mythril & mythril/myth \\
                Slither & github.com/crytic/slither & trailofbits/eth-security-toolbox \\
                Securify & github.com/eth-sri/securify2 & Dockerfile \\
                Smartian & github.com/SoftSec-KAIST/Smartian & Dockerfile \\
                Sailfish & github.com/ucsb-seclab/sailfish & holmessherlock/sailfish:latest \\
                eTainter & github.com/DependableSystemsLab/eTainter & Manual construction \\
                Echidna & github.com/crytic/echidna & ghcr.io/crytic/echidna/echidna \\
                \hline
        \end{tabular}
        }
        \label{tab:selected_tools}
\end{table}

\subsection{Experiment}

\subsubsection{Setup}

Specifically, we directly downloaded the Docker images of Mythril, Slither and Sailfish from Dockerhub~\cite{dockerhub}. For Securify and Smartian, we built their Docker images according to the Dockerfiles in their GitHub repositories. 

Based on Docker images, we used the selected tools to analyze smart contracts in \textsc{DAppSCAN-Bytecode}. Specifically, according to the input of the tools, we perform Mythril, Slither, Securify and Sailfish on the Solidity code of smart contracts, and perform Smartian and eTainter on the Bytecode of smart contracts. To determine the time budget of the tools, we randomly sampled 364 smart contracts (a statistically significant sample size considering a confidence level of 95\% and a confidence interval of 5\%) from the \textsc{DAppSCAN-Bytecode} dataset and evaluated the execution time of the tools. Then we found that 90.3\% ($\frac{1972}{364*6}$) analysis tasks could be finished within five minutes. Thus, we set an execution time budget of five minutes for each smart contract analysis task. If the time budget is up, we stop the execution and export the analysis results. For fair evaluation, we ran the selected tools with a default configuration similar to the setting followed by Durieux et al.~\cite{durieux2020empirical}. All experiments were conducted on a Ubuntu machine with an Intel Core i9-10980XE CPU (36 cores and 72 threads) and 256 GB of memory.

\subsubsection{Result Analysis}

We evaluate the effectiveness of the tools based on the automated analysis results and provide some insightful findings.

\begin{figure} 
	\begin{center}
		\includegraphics[width=0.85\linewidth ]{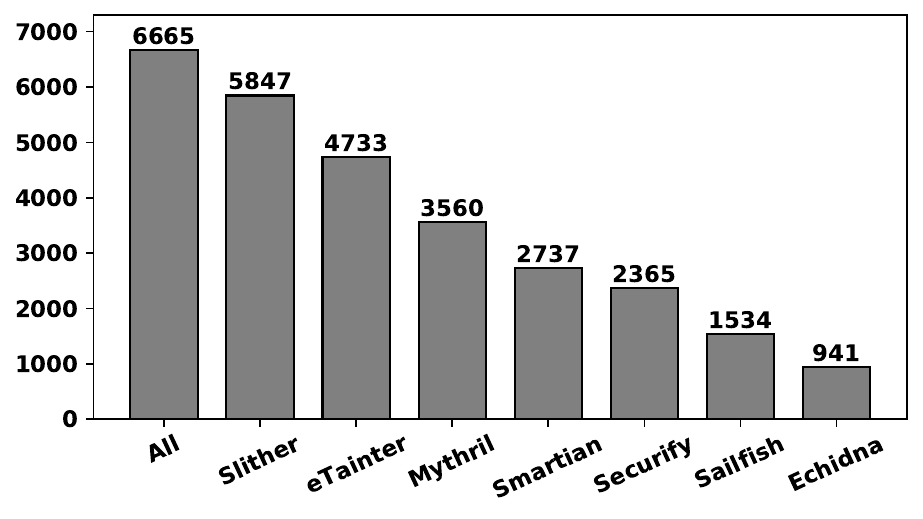} 
		\caption {The number of smart contracts being successfully analyzed by each tool.} 
		\label{Fig:success_analysis} 
	\end{center} 
\end{figure}


\noindent\textbf{Successful Analysis Rate.} Fig.~\ref{Fig:success_analysis} shows the number of smart contracts that are successfully analyzed by each tool. We notice that Slither obtains the highest successful analysis rate, about 88\% ($\frac{5,847}{6,665}$), this is likely because Slither is a relatively mature industrial product with continuous updates, while other academic tools receive little upgrade once proposed. We also find that other tools fail to analyze many smart contracts. We perform further analysis and illustrate the reasons as follows: (1) The analysis scopes of tools are limited by Solidity versions. Specifically, although we have selected the tools that support analyzing smart contracts with multiple versions, there are still some versions that are not supported by tools (e.g., the newly proposed versions after the tools are proposed). (2) Some design flaws in tools lead to failure of analysis, such as the tool running out of memory during analysis and throwing error, the tool failing to compile contracts with its built-in compilation settings.

\begin{table*}[t] 
    \centering
    \small
        \caption{Representative vulnerability detection tools used in our study. "All" means the number of all vulnerable Solidity files. "/" means the tool can not detect the corresponding weakness. "Union" means the number of the union of all detected vulnerable Solidity files by tools.}
        \resizebox{0.95\linewidth}{!}{
            \begin{tabular}{l|lll|lll|lll|lll|lll|lll|lll|c|l}
                \hline
                \multirow{2}{*}{\textbf{ID}} & \multicolumn{3}{c|}{\textbf{Mythril}} & \multicolumn{3}{c|}{\textbf{Securify}} & \multicolumn{3}{c|}{\textbf{Slither}} & \multicolumn{3}{c|}{\textbf{Smartian}} & \multicolumn{3}{c|}{\textbf{Sailfish}} & \multicolumn{3}{c|}{\textbf{eTainter}} & \multicolumn{3}{c|}{\textbf{Echidna}} & \textbf{Union} & \multirow{2}{*}{\textbf{All}} \\
                & TP & FP & FN & TP & FP & FN & TP & FP & FN & TP & FP & FN & TP & FP & FN & TP & FP & FN & TP & FP & FN & TP & \\
                \hline
101  & 2 & 71 & 90 & / & / & / & / & / & / & 0 & 105 & 92 & / & / & / & / & / & / & 1 & 18 & 91 & 3 & 92  \\
103  & / & / & / & 37 & 1298 & 53 & 82 & 4132 & 8 & / & / & / & / & / & / & / & / & / & / & / & / & 82 & 90  \\
104  & 0 & 8 & 60 & 0 & 114 & 60 & 13 & 538 & 47 & 0 & 26 & 60 & / & / & / & / & / & / & / & / & / & 13 & 60  \\
105  & / & / & / & 0 & 6 & 14 & 0 & 56 & 14 & 0 & 4 & 14 & / & / & / & / & / & / & / & / & / & 0 & 14  \\
106  & 0 & 20 & 1 & 0 & 8 & 1 & 0 & 15 & 1 & 0 & 12 & 1 & / & / & / & / & / & / & / & / & / & 0 & 1  \\
107  & 15 & 115 & 68 & 1 & 13 & 82 & 45 & 459 & 38 & 0 & 0 & 83 & 0 & 31 & 83 & / & / & / & / & / & / & 46 & 83  \\
108  & / & / & / & 0 & 124 & 4 & / & / & / & / & / & / & / & / & / & / & / & / & / & / & / & 0 & 4  \\
109  & / & / & / & 0 & 322 & 1 & 0 & 59 & 1 & / & / & / & / & / & / & / & / & / & / & / & / & 0 & 1  \\
110  & 0 & 106 & 0 & / & / & / & / & / & / & 0 & 350 & 0 & / & / & / & / & / & / & / & / & / & 0 & 0  \\
112  & 0 & 8 & 6 & 0 & 1 & 6 & 0 & 19 & 6 & / & / & / & / & / & / & / & / & / & / & / & / & 0 & 6  \\
113  & 0 & 26 & 7 & / & / & / & / & / & / & 0 & 4 & 7 & / & / & / & 0 & 0 & 7 & / & / & / & 0 & 7  \\
114  & / & / & / & 1 & 37 & 41 & / & / & / & / & / & / & 1 & 56 & 41 & / & / & / & / & / & / & 2 & 42  \\
115  & 0 & 10 & 1 & / & / & / & 1 & 1 & 0 & 0 & 2 & 1 & / & / & / & / & / & / & / & / & / & 1 & 1  \\
116  & 0 & 126 & 3 & 0 & 28 & 3 & 0 & 305 & 3 & 0 & 6 & 3 & / & / & / & / & / & / & / & / & / & 0 & 3  \\
119  & / & / & / & 0 & 40 & 0 & 0 & 331 & 0 & / & / & / & / & / & / & / & / & / & / & / & / & 0 & 0  \\
120  & / & / & / & / & / & / & 0 & 24 & 5 & / & / & / & / & / & / & / & / & / & / & / & / & 0 & 5  \\
124  & 0 & 1 & 0 & 0 & 248 & 0 & / & / & / & 0 & 5 & 0 & / & / & / & / & / & / & / & / & / & 0 & 0  \\
125  & / & / & / & / & / & / & 0 & 72 & 1 & / & / & / & / & / & / & / & / & / & / & / & / & 0 & 1  \\
128  & / & / & / & / & / & / & 1 & 167 & 48 & / & / & / & / & / & / & 0 & 10 & 49 & / & / & / & 1 & 49  \\
131  & / & / & / & 0 & 232 & 44 & 2 & 222 & 42 & / & / & / & / & / & / & / & / & / & / & / & / & 2 & 44  \\
135  & / & / & / & / & / & / & 105 & 2735 & 30 & / & / & / & / & / & / & / & / & / & / & / & / & 105 & 135  \\
\hline
100  & / & / & / & / & / & / & / & / & / & / & / & / & / & / & / & / & / & / & / & / & / & / & 6  \\
102  & / & / & / & / & / & / & / & / & / & / & / & / & / & / & / & / & / & / & / & / & / & / & 19  \\
117  & / & / & / & / & / & / & / & / & / & / & / & / & / & / & / & / & / & / & / & / & / & / & 1  \\
121  & / & / & / & / & / & / & / & / & / & / & / & / & / & / & / & / & / & / & / & / & / & / & 1  \\
122  & / & / & / & / & / & / & / & / & / & / & / & / & / & / & / & / & / & / & / & / & / & / & 2  \\
126  & / & / & / & / & / & / & / & / & / & / & / & / & / & / & / & / & / & / & / & / & / & / & 1  \\
133  & / & / & / & / & / & / & / & / & / & / & / & / & / & / & / & / & / & / & / & / & / & / & 2  \\

                \hline
        \end{tabular}
        }
        \label{tab:analysis_results}
\end{table*}

\noindent\textbf{Weakness Detection.} Based on the labels in our dataset, we perform a file-level evaluation on the tools' analysis results. Specifically, if the tools report the weaknesses in the vulnerable Solidity files with corresponding labels, we consider that the tools successfully detect the true weaknesses. Table~\ref{tab:analysis_results} shows the results of the analysis, which only shows the SWC weaknesses whose number is greater than 0 and can be detected by at least one tool. In this table, if the element in the table is represented as a number, it means that the tool for the corresponding column can detect the SWC-ID weakness of the corresponding row. In total, the tools support the detection of 21 kinds of SWC weakness. Slither supports detecting the most kinds of SWC weakness, and some tools (e.g., Sailfish and eTainter) only support detecting a few kinds of SWC weakness.

In general, the tools show their limited effectiveness in detecting SWC weaknesses within audit reports. Next, we illustrate the detection results in detail from three perspectives: True Positive, False Positive, and False Negative.

\textbf{True Positive.} For the 21 kinds of SWC weakness that the tools support detection, we notice that the tools only successfully detect a few true SWC weaknesses in audit reports (see the "Union" column), most of which are reported by Slither. The three most weaknesses detected by the tools are SWC-103 (Floating Pragma), SWC-135 (Code With No Effect), and SWC-107 (Reentrancy) with successful detection rates of 91.1\% ($\frac{82}{90}$), 77.8\% ($\frac{105}{135}$), and 55.4\% ($\frac{46}{83}$), respectively. 

\textbf{False Positive.} Overall, the tools report a huge number of false positive weaknesses. Specifically, some kinds of SWC weakness such as SWC-103 (Floating Pragma), contain more than 1,000 false positives reported by tools. The limited performance (e.g., inaccurate detection mechanism) of tools mainly results in false positives. For example, reentrancy could be prevented by \textit{Reentrancy Lock}, which is not considered by tools and leads to false positives~\cite{zheng2023turn}. In particular, this result is similar to work~\cite{perez2021smart}, which found the high false positive rate of tools.

\textbf{False Negative.} For some kinds of SWC weakness, the tools show their high false negative rates. In particular, some kinds of SWC weakness are commonly seen in audit reports but only few of them are successfully detected by tools, including SWC-104 (Unchecked Call Return Value), SWC-114 (Transaction Order Dependence), SWC-128 (DoS With Block Gas Limit) and SWC-131 (Presence of Unused Variables). These results demonstrate the limitations of tools to protect real-world DApps from weaknesses.


\textbf{Unsupported Weaknesses.} As shown in Table~\ref{tab:analysis_results}, 7 kinds of SWC weakness in audit reports are not supported detecting by any tool. Among them, SWC-102 (Outdated Compiler Version) is the most common weakness in audit reports, while other kinds of SWC weakness are small in number. We perform further analysis on these kinds of SWC weakness and find that they are difficult to design detection patterns. For example, the Solidity compiler is updated frequently, making it hard for tools to follow the latest compiler and detect SWC-102 weaknesses.

%% file: S7-Discussion.tex
\section{Discussion}
\label{sec:discussion}
\subsection{Implications}

\subsubsection{Limitations of Current Academic Tools}
In Section~\ref{sec:motivation}, we list 20 academic tools that could detect SWC weaknesses. We tried to use these tools on the \textsc{DAppSCAN-Source} dataset proposed in Section~\ref{sec:source_dataset}. However, we found that none of these academic tools was able to support the analysis of raw DApp projects. The key reason for this failure was that most DApps were developed using a smart contract framework, such as Truffle~\cite{Truffle}, Brownie~\cite{Brownie}, or Hardhat~\cite{Hardhat}, Embark~\cite{Embark}. There are {\averageSolNum} Solidity files on average in the DApps of our dataset. Smart contracts can inherit from other contracts, and the file structure of DApp projects is usually complicated due to the import relationship (for details, see Section~\ref{sec:background_dapp}). We found that only Slither~\cite{feist2019slither}, a tool from the industrial team named TrailofBits~\cite{tob} supported the analysis of DApps developed using smart contract frameworks. This finding may also hint that academic work should focus more on the real-world development process.

\subsubsection{Automatic Software Engineering for Smart Contracts}


In this paper, we have proposed a dataset that contains {\SWCNum} SWC weaknesses from {\DAppNum} DApps. The scale of our dataset can be used to perform some automatic software engineering research. For example, about 86.2\% of the DApps in our dataset are projects on Github or Gitlab. For each SWC weakness, we have provided the location of the vulnerable code, which can be used to automatically locate/predict weaknesses. In addition, the fixed code recorded within the associated Github commit in audit reports can be utilized to develop code repair technologies. Specifically, we will further update and complete the dataset in our future work, such as attaching the fixing commit directly in the dataset and searching for more audit reports/real-world projects.

\subsubsection{Non-SWC Weaknesses in Audit Reports}

Due to the limited labor force, we only labeled codes with SWC weaknesses, as the SWC Registry covered a wide range of smart contract issues and is commonly used in existing research (as shown in Table~\ref{tab:tools}). However, 82.3\% ($\frac{7,536}{\allWeaknessNum}$) of the weaknesses in the audit reports of our dataset are not included in the SWC Registry. 

To study non-SWC weaknesses, we sampled and analyzed 366 weaknesses, a statistically significant sample size considering a confidence level of 95\% and a confidence interval of 5\% referring to~\cite{zheng2023turn}. After manual analysis, we summarize their main conditions as follows: (1) The reported items are not related to the security of smart contracts, such as the gas optimization and the code standardization suggestions; (2) The reported weaknesses are functional bugs or design vulnerabilities related to the specific business of each DApp, which cannot be broadly categorized within the SWC registry. For example, a token transfer logic is incorrectly implemented in the deposit function, which enables attacker to obtain profits unfairly~\cite{tokenTransfer}. Note that such a condition is common in non-SWC weaknesses, underscoring their significant relevance to the advancement of security within the DeFi ecosystem. In general, both are hard to classify with general rules. Furthermore, SWC registry has not been updated since 2020, so SWC registry cannot cover these new and non-classical weaknesses.


Given that the majority of items in audit reports are non-SWC weaknesses, we also provide descriptions and related audit reports in our dataset. In future work, we will conduct an in-depth empirical study on these weaknesses and summarize typical and common issues to facilitate further research, especially for functional bugs.


\subsection{Threats to Validity}

\subsubsection{Internal Validity} 

In this study, we proposed two types of datasets, i.e., \textsc{DAppScan-Source} and \textsc{DAppScan-Bytecode}. Both of them required significant labor, which may involve some errors. Specifically, in \textsc{DAppScan-Source}, we manually analyzed {\ReportCodeNum} audit reports and identified SWC weakness-related information from them. In\textsc{DAppScan-Bytecode}, we manually collected the missing third-party libraries. To ensure accuracy, all manual processes were performed by at least two experienced researchers. Additionally, the dataset is open-source on GitHub. \textbf{If users identify any errors, they can submit issue reports through Github, and we will update our dataset promptly.}

\subsubsection{External Validity} 

The Ethereum ecosystem is rapidly evolving. For example, Solidity v0.6.1 was released in January 2020, while the latest version as of April 2023 is Solidity v0.8.20. There are many differences between the various compiler versions, and some kinds of SWC weakness may disappear due to updates of the Solidity compiler. Although developers are allowed to choose older Solidity compiler versions, there is a trend towards the use of newer compilers. Fortunately, most of the DApps we collected in this paper used compiler versions greater than 0.6+. In the future, we will update our dataset to ensure that it remains up to date.

%% file: S8-Related.tex
\section{Related Work}
\label{sec:related}
Recently, many smart contract analysis tools have been developed, and several studies have investigated their efficacy. 


Durieux et al.~\cite{durieux2020empirical} conducted an empirical study based on nine automated smart contract analysis tools. To compare and reproduce their research, they create two datasets: (1) 69 annotated vulnerable smart contracts with 115 vulnerabilities (97 of them were tagged as one of the DASP vulnerabilities) that were used to evaluate the precision of tools; and (2) 47,518 open-source smart contracts collected from Etherscan to compare different tools. According to their results, only 29 out of 69 vulnerabilities were detected by the tools. Among the nine tools, Mythril obtained the highest accuracy, but this was only 27\%. The nine tools flagged 97\% of smart contracts in the large-scale dataset as vulnerable, which shows the high positives of current tools. 

Perez et al.~\cite{perez2021smart} evaluated whether vulnerable smart contracts detected by academic tools can be actually exploited. First, they used six academic tools to analyze 821,219 contracts, and 23,327 of them were flagged as vulnerable by at least one tool. Then, the authors performed bytecode-level analysis to confirm potential exploits. They found that only 463 (1.98\%) of them had been exploited since their deployment. The balances held in these contracts were 8,487 ETH (0.27\%), whereas the total balance in the 23,327 vulnerable contracts was 3 million ETH. These results reflect that the most vulnerabilities reported in academic tools are either false positives or not exploitable.

Yi et al.~\cite{yi2021diving} built a blockchain vulnerability dataset, which contained three levels, i.e., file-level, text-level, and code-level vulnerabilities. To build the dataset, they collected commit messages, issue reports, and pull requests from Github. Then, they conducted an empirical study to investigate common blockchain vulnerability types, blockchain-specific patch code patterns, and susceptible blockchain modules. Their results showed that modules related to consensus, wallet, and networking were more vulnerable than other modules; and about 70\% vulnerabilities on blockchain systems were similar to those in traditional projects. The remaining 30\% of the vulnerabilities were classified into 21\% blockchain-specific vulnerability patterns. 

Zhang et al.~\cite{zhang2023demystifying} systematically investigated 516 unique real-world smart contract vulnerabilities from \textit{Code4rena}~\cite{code4rena} contests and real-world exploit reports, 74.6\% of which are exploitable. They further studied the scope of detectable vulnerabilities and found that existing tools could only deal with about 20.5\% of real-world exploitable vulnerabilities. For the remaining undetectable vulnerabilities, they categorized them into seven types and performed in-depth analysis, such as root causes and distributions.

Di et al.~\cite{di2023evolution} performed a large-scale tool evaluation on 248,328 distinct Ethereum smart contracts. They utilized the open-source framework Smartbugs~\cite{ferreira2020smartbugs} to conduct the automated analysis on the bytecode of smart contracts. Consequently, while the tools reported a total of 1,307,686 potential weaknesses, over time (block height) the number of reported vulnerabilities is decreasing and the tools are degrading to varying degrees.

Chaliasos et al.~\cite{chaliasos2023smart} evaluated five state-of-the-art automated security tools based on 127 high-impact real-world attacks. They also surveyed 49 developers and auditors who work on leading DeFi protocols. Their findings indicated that the tools could only detect about 8\% of the attacks, which are all related to reentrancy vulnerabilities. Existing tools do not adequately address logic-related bugs and vulnerabilities in the protocol layer, which are viewed by practitioners as significant threats.

\textbf{Differences to our work: } Compared to other works, our work provided a larger-scale and labeled dataset containing {\allWeaknessNum} weaknesses with {\SWCNum} SWC weaknesses in the form of DApp projects (rather than individual contracts as previous works). In particular, we also proposed a tool that could automatically compile DApp projects and provided a compilable-contract dataset to facilitate contract analysis. Furthermore, we conducted extensive experiments and highlighted the limitations of current tools for analyzing real-world smart contracts in DApps.


%% file: S9-Conclusion.tex
\section{Conclusion and Future Works}
\label{sec:conclusion}

In this paper, we proposed two large-scale datasets with SWC weaknesses to facilitate the evaluation of smart contract analysis tools. To construct the first \textsc{DAPPSCAN-Source} dataset, we recruited participants to analyze {\ReportCodeNum} audit reports. Our dataset contains a total of {\allWeaknessNum} weaknesses, with {\SWCNum} highlighted as SWC weaknesses, providing their descriptions and locations in the source code. The SWC weaknesses in our dataset originated from {\DAppNum} DApps. On average, each DApp had {\averageSolNum} Solidity smart contract files with {\averageLOC} lines of code, and 66.3\% of DApps used compiler versions higher than 0.6. The second dataset, \textsc{DAPPSCAN-Bytecode}, is compiled from the \textsc{DAPPSCAN-Source} dataset using a tool that automatically parses the dependency relations of smart contracts within DApps. This dataset comprises {\compiledByteNum} smart contracts with {\SWCByteNum} SWC weaknesses. Finally, based on the second dataset, we conducted an empirical study to assess the effectiveness of seven state-of-the-art tools in SWC weakness detection. The results reveal poor performance in both effectiveness and successful detection rate, highlighting the need for tools capable of detecting real-world DApp smart contracts.


In the future, we plan to build a website to display the results of this paper, such as the frequency and security events of SWC weaknesses. The website will facilitate the search and downloading of weakness cases. Moreover, we aim to include additional audit reports provided by security teams and will collect and analyze these reports regularly to expand our dataset. In addition, we also aim to study the non-SWC weakness and summarize common issues to facilitate future research.
